\documentclass[final,english]{bullsrsl}[2022/06/15]

\newcommand{\kms}{km\,${\rm s}^{-1}$}

\usepackage[latin1]{inputenc}
\usepackage[T1]{fontenc}

\usepackage[numbers]{natbib} 
\usepackage{graphicx}

\begin{document}
\title{Hot and cool: Characterising the companions of red supergiant stars in binary systems}

\author[affil={1}, corresponding]{Lee R.}{Patrick}
\author[affil={2}]{Ignacio}{Negueruela}
\affiliation[1]{Centro de Astrobiolog\'ia (CSIC-INTA), Ctra.\ Torrej\'on a Ajalvir km 4, 28850 Torrej\'on de Ardoz, Spain}
\affiliation[2]{Departamento de F\'{\i}sica Aplicada, Universidad de Alicante, E-03690 San Vicente del Raspeig, Alicante, Spain}
\correspondance{lrpatrick@cab.inta-csic.es}
\date{13th October 2020}
\maketitle


%

\begin{abstract}
In this article we study the nature of the recently identified populations of hot companions to red supergiant stars (RSGs).
To this end, we compile the literature on the most well characterised systems with the aim of better understanding the hot companions identified with ultra-violet photometry and confirmed with Hubble Space Telescope spectra in the Local Group.
We identify 9 systems with current masses greater than around 8\,M$_\odot$ that have constraints on their orbital periods, which are in the range 3 to 75\:a. 
Antares ($\alpha$~Sco) is the obvious outlier in this distribution, having an estimated orbital period of around 2000\:a.
Mass-ratio ($q=M_2/M_{\rm RSG}$) estimates are available only for 5 of the compiled systems and range between $0.16 < q < 1$.
Ongoing efforts to identify and characterise hot companions to RSGs in the SMC have revealed 88 hot companions that have observational constraints free of contamination from the RSG component. 
We present a summary of a recently conducted HST UV spectroscopic survey that aims to characterise a subset of these companions.  
These companions show a flat-$q$ distribution in the range $0.3 < q < 1$.



\end{abstract}


\msccodes{65F15 65G50 15-04 15B99}






\section{Introduction}

The majority of massive stars are born in binary or higher-order multiple systems~\citep{2014ApJ...780..117S,2014ApJS..213...34K,MdS17,Offner_2023}, with $\sim$70\% expected to
interact during their lifetimes~\citep{2012Sci...337..444S}.
These interactions have profound effects on the evolution of the stars in such systems~\citep{2013ApJ...764..166D} and change the landscape of expected  supernova explosions~\citep{1992ApJ...391..246P, 2017PASA...34....1D}, impacting on the formation of stellar mass double compact object (DCO) binaries~\citep{2017A&A...604A..55M}.
The first steps are already being taken to examine how multiplicity affects the evolution of stellar populations~\citep{2008MNRAS.384.1109E,2020ApJ...888L..12W}. Such simulations are also beginning to produce estimates of the binary properties of their evolved products that include DCOs~\citep{2020A&A...638A..39L}.

As the final phase of evolution for the majority of massive stars, the red supergiant (RSG) stars are a vital piece of the stellar multiplicity puzzle. 
The numbers, luminosity and effective temperatures of populations of RSGs are important for constraining stellar physics but the contribution of binary evolution to this picture is poorly constrained. 
For example, binary merger products are observed to be common in the RSG phase~\citep{2019MNRAS.486..266B,2019A&A...624A.128B}, with up to 50\,\% of RSGs thought to be `red stragglers'~\citep{2019A&A...624A.128B,2020A&A...635A..29P}.
Moreover, mass-transfer that initiates during the RSG phase results in partial envelope stripping and contributions to the diversity of observed supernova. 
In addition, RSG binary systems that survive a supernova explosion may result in the formation of DCOs~\citep{2018MNRAS.481.1908K,2020A&A...638A..39L}.


Despite being important for multiple stellar endpoints, there are very few examples of well characterised RSG binaries.
As the known population of RSG binaries is dramatically expanding both in the Galaxy and in the Local Group, it is vital to anchor these newly discovered systems to the best studied RSG binaries.
In this respect, the known population of RSG binary systems is vital for guiding expectations and theoretical simulations of the types of orbital configurations and companions.
In this article, we first provide a brief theoretical perspective on the constraints that prior evolution places on the orbital configurations for RSGs in Section~2.
We then review the binary systems with well determined orbital parameters and outline the techniques that have and are being employed to identify and characterise RSG binary systems in the Local Universe. 
With this information in hand, in Section~4 we turn our attentions to the recently identified population of RSG binaries in the Local Group of galaxies and focus ultra-violet (UV) studies of RSG binaries in the Milky Clouds (MCs). 
We summarise this article in Section~5.



\section{Theoretical expectations}

RSGs are among the physically largest stars in the Universe.
As stars evolve off the main-sequence, their radii dramatically expand on a thermal timescale.
As the majority of massive stars within the 8--40\,M$_\odot$ range are born with a close companion~\citep{2012Sci...337..444S,BLOeM-BV,MdS17,Offner_2023}, it is likely that when the more massive component in a binary system expands, a mass-transfer interaction occurs, the result of which is a hot stripped star~\citep{2008MNRAS.384.1109E} and not a RSG.
In this theoretical framework, the only systems that survive to become an observable RSG binary system are those that have had no prior mass-transfer or systems that have merged while on the main-sequence and subsequently evolved to the RSG phase.
If we assume that the current size of the star determines the minimum orbital period allowed, we may set a theoretical limit for observational searches for RSG binary systems. 
Figure~1 shows the result of simulations of 10,000 binary systems at each grid point for a range of orbital periods and $q$ values with a random inclination angle with respect to the observer, primary masses in the range $8 <$M/M$_\odot < 40$ and eccentricity values drawn from a probability distribution $P(e) \sim e^{+0.8}$ within the range $0 < e < 0.8$, which is appropriate for long orbital period massive stars according to~\citet{2017ApJS..230...15M}. 
The underlying black solid contour lines in Figure~1  show lines of fixed semi-amplitude velocity ($K_1$).
In this figure, dashed coloured lines show the minimum orbital period possible as a function of $q$ for three case study theoretical RSGs.
This figure shows that orbital periods around 1000\,days are excluded and the orbital systems that are allowed have $K_1$ values in the range 5--25\kms.
To tether this to observations, we calculate the velocity change in one year for a representative system.
A system with $\log\,P = 3.0$ and $q=1.0$ has $K_1\sim25\:$\kms, which over the course of 1 year results in a velocity change of $\sim10\:$\kms.

VV~Cep has an orbital period of $\log P= 3.9$, $K_1 = 20$\,\kms\ and $q=1.0$, which stretches the limits of these simulations, but we caution that these distributions are averaged over both an eccentricity distribution and a primary mass distribution and because of that individual systems may be outside of these predicted curves. 
Simulating VV~Cep specifically the simulations are able to reproduce the observed $K_1$.

\begin{figure}
\centering
\includegraphics{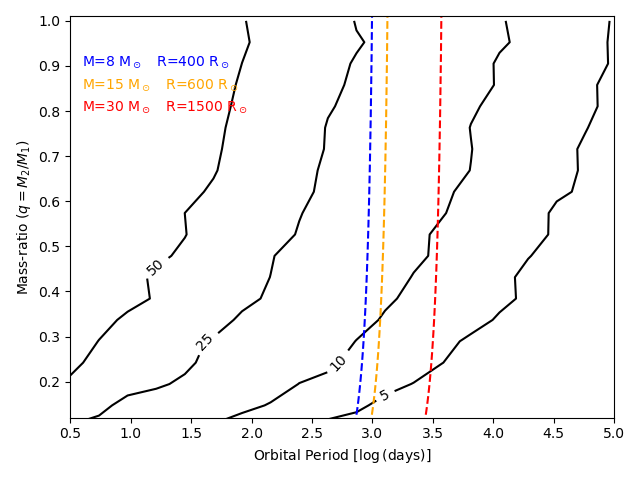}
\bigskip

\begin{minipage}{12cm}
\caption{Orbital period -- mass-ratio ($q=M_2/M_1$) diagram for a set of simulated binary systems covering orbital periods in the range 3 to 10$^5$\,$d$. The minimum allowed orbital period as a function of $q$ for three RSG case studies at 8, 15 and 30\,M$_\odot$ with representative radii. Black solid lines show lines of constant semi-amplitude velocity for the simulated systems on the 5, 10, 25 and 50 \kms\ levels.}
\end{minipage}
\end{figure}

The types of companions that are theoretically expected for RSGs has not been fully explored to date, however, simulations show that the most common companion is expected to be main-sequence stars \citep[C. Sch\"urmann \textit{private communication};][]{2020ApJ...900..118N}).
Other companions that are expected in smaller numbers are neutron stars, black holes and white dwarf stars.






\section{Constraints on the orbital configurations of RSGs in binary systems}
In this section, we describe RSG binaries that have known orbital periods.
This, in turn, will allow a more accurate interpretation of whether or not the newly observed systems in external galaxies fit in with expectations.
We exclude from this discussion stars which have RSG masses below $\sim$8\,M$_\odot$, thereby omitting many of the systems that have been considered RSG binary systems in the literature, such as 31~Cyg~\citep[M$_{\rm RSG}=6.7\:$M$_\odot$,][]{2018csss.confE..46B} and $\zeta$~Aur~\citep[M$_{\rm RSG}=5.8\:$M$_\odot$,][]{1995ApJ...455..317B}.
This choice assumes that the current mass reflects the initial mass of the star, which is, in effect, assuming that mass-transfer has not taken place. 
The validity of this assumption is difficult to assess. On the one hand, by their very nature, such large orbital periods suggest no significant mass-transfer events; on the other hand, at least two systems (32~Cyg and V766~Cen) have orbital periods that suggest mass-transfer may have occurred.
Although we caution that RSG masses in the Galaxy are difficult to accurate constrain.

Identification of binary systems in general focuses on identifying anomalies in the observations of stars that cannot be attributed to anything other than the presence of an additional source.  
As RSGs are particularly bright and variable, this has presented a significant obstacle to the detection of other sources in their vicinity.
Most RSG binary systems in the Galaxy have been identified via spectral signatures that are not expected from a single RSGs. 
As described below, the companion of Antares was identified visually using lunar occultations in the 1800s, whereas the realisation that V766~Cen is a RSG binary system has come only recently. 
With the exception of V766~Cen and Antares, the remaining stars are often termed VV~Cep variables, which refers to a RSG with a hot companion. 
Here we generalise that classification and refer to RSG binaries rather than VV~Cep systems. 
\citet{2020RNAAS...4...12P} compiled a list of stars that have been spectroscopically identified as a RSG binary system from the literature.

Observations at ultra-violet wavelengths have long since been realised to be an effective wavelength regime to study RSG binary systems based on the contrasting light ratios between the two components and much work with the IUE satellite focused on how the hot companion illuminates the cool star atmosphere~\citep[see][]{2015ASSL..408.....A}. 

Multi-epoch spectroscopic techniques are effective tools to study binary motion and this has been used in multiple studies to identify the presence of RSGs in binary systems~\citep{patrick2019, 2020A&A...635A..29P, 2021MNRAS.502.4890D}. However, such studies typically cannot go beyond inferring the presence of binaries, rather than characterising their orbital configurations. 
In their study of RSG variability, \citet{2007A&A...469..671J} identified four RSG binary systems in their sample and subsequently excluded them from further analysis. Three of their four systems (HD~203338, XX~Per and U~Lac; the other being VV~Cep) have no published orbital period and, as such, the spectroscopic observations that were excluded from~\citet{2007A&A...469..671J} may yet prove constraining on an orbital solution.
Going beyond this, it is likely that many of the RSG binary systems listed in~\citet{2020RNAAS...4...12P} have sufficient historical observations to estimate orbital periods.

Table~1 lists the well studied RSG binary systems along with some basic orbital and stellar parameters.
Where quantities are uncertain, we list ranges quoted in the literature.
Each system is discussed individually below where references for the values quoted in the table are provided, whenever possible.

\begin{table}
\centering
\begin{minipage}{88mm}
\caption{Orbital configurations of RSG binary systems with constraints on orbital solutions.}
\end{minipage}
\bigskip

\begin{tabular}{lll cccc}
\hline
\textbf{ID} & \textbf{Alt. ID} & \textbf{Components} & \textbf{Period} & \textbf{$\log L_{\rm RSG}/L_{\odot}$} & \textbf{M$_{\mathrm{RSG}}$}  & \textbf{$q=$}\\
& & & & [dex] & [\textbf{M$_{\odot}$}] & $M_2/M_1$\\

\hline
32 Cyg & HR~7751  & K5\,Ib + B6\,V & 1148\,d &3.82\,$\pm$\,0.08 & 7.5--10 & 0.5 \\
V766~Cen  & HR~5171~A & RSG + YSG & 1304\,d & 5.8\,$\pm$\,0.4& 27--36 & 0.16$^{+0.4}_{-0.07}$\\
AZ~Cas & -- & K5\,Iab-Ib + B & 3406\,d & 4.2 & -- & -- \\
VV~Cep & -- & M2\,Iab + B0-2\,V  & 7431\,d & 5.3\,$\pm$\,0.2 & 20 &  $\sim$1 \\
KQ Pup & HD60414 & M1\,Iab + B2 & 9752\,d & 4.6& 13--20 & $\sim$1  \\
& Boss~1985\\
HD~42474 & WY Gem & M1\,Ib + B2 & $>40$\:a &  4.6 & 20 & 0.75 -- 1 \\
5 Lac & HD 213310 & K6-M0\,I + B7/8 & 41.95\,a & 4.4 & 5.1 & -- \\ 
HD~203338 & V381 Cep & M1\,Ib + B2 & $\sim$ 75\,a & 5.2 & -- \\
 & Boss 5481\\
$\alpha$ Sco & Antares & M1.5\,Iab + B2.5\,V & 2700\,a & 5.0 & 18 & 0.4 \\
\hline
\end{tabular}
\end{table}

\noindent\textbf{VV Cep}\\
The orbital configuration of VV~Cep is best described in the text book \textit{Giants of the Eclipse}~\citep{2015ASSL..408.....A}. 
By some distance VV~Cep is the most well studied RSG binary system and the orbital parameters determined for this system are generally considered precise.
Photometric and spectroscopic monitoring have covered multiple orbits and eclipse monitoring continued throughout the recent primary eclipse~\citep{2020JAVSO..48..118P,2022JAVSO..50..264P}.
The fact that the orbital parameters have changed little since the pioneering work of~\citet{1977JRASC..71..152W} illustrates the well defined nature of the orbital parameters.

The B-type companion is embedded in an accretion and shock nebula and the stellar spectrum is (almost) never directly observed even in the UV.
Debate continues over whether or not this system has experienced mass transfer via Roche-lobe overflow.
In general, mass-transfer acts to circularise orbits and an eccentric orbit is observed for this system.
However, clear evidence of interaction between the two components exists in the shrouding of the hot companion.
These two pieces of evidence are reconciled by the assumption that interaction between the wind of the RSG and the hot companion has taken place rather than mass-transfer.
Although, we note here that a non-circular orbit does not rule out previous mass-transfer.

Given that the B-type star is never directly observed, the RV information from the enshrouded companion is exclusively determined from the H$\alpha$ profile. 
For this system, the H$\alpha$ profile is a complex, double-peaked emission structure which is variable.
The central emission peak of the H$\alpha$ profile - which was presumably determined via a by-eye analysis in~\citet{1977JRASC..71..152W} - is assumed to track the motion of the hot companion. 

\noindent\textbf{V766 Cen}\\
V766~Cen (= HR~5171A) was considered to be one of the few examples of a yellow hypergiant (YHG) star in the Galaxy~\citep{1971ApJ...167L..35H}, with spectral type G8\,Ia$^{+}$ or K0\,Ia$^{+}$ (the spectral type may be genuinely variable; see \citep{2014A&A...563A..71C} for references). 
In addition, it has a known supergiant visual companion with a spectral type of B0\,Ib ~\citep{1971ApJ...167L..35H}.
\citet{2014A&A...563A..71C} discovered that the central component, the YHG, was an eclipsing binary system.
With VLTI spectro-interferometry, \citet{2017A&A...597A...9W} refined the stellar parameters of the primary source and argued that, given its huge size and the presence of strong CO bandheads, the star is better classified as a high-luminosity RSG rather than a YHG, although we note that the effective temperature determined has a particularly large uncertainty~\citep{2017A&A...597A...9W}.
\citet{2017A&A...606L...1W} provided additional evidence with the VLTI that the central component is indeed a binary system and estimated the mass of the RSG component to be between 27 -- 36\,M$_\odot$ and the companion to be 5$^{+15}_{-3}$\,M$_\odot$.
This system presents two unique qualities in so much that it is likely among the most massive RSGs known and it is the only RSG binary system known to have a potential RSG companion. 

The distance to HR 5171 has been traditionally constrained by its apparent association with the H\,{\sc ii} region Gum~48d, on top of which it appears projected. Literature values range from 3.1 to 3.6~kpc \citep{karr09}, with the latter value generally adopted. Nevertheless, \textit{Gaia} DR3 data favours a lower distance for the stars in this area. The stellar aggregate HSC~2611 \citep{hunt24}, which likely represents the emerged early-type population in the area, has a distance of 3.0~kpc, and the parallax of HR~5171B is fully compatible with this value. 

Despite the recent classification as a RSG, the spectral type of V766~Cen is a late G-type hypergiant~\citep{1971ApJ...167L..35H,1973MNRAS.161..427W}. 
There is therefore a clear mismatch between the determined stellar parameters of the central source and the spectral classification.
The red dashed line in Figure~1 corresponds to the stellar parameters of V766~Cen as determined by~\citet{2017A&A...597A...9W}.
With these stellar parameters the minimum orbital period for a non-interacting configuration is larger than 3000\,d, which either argues for a smaller RSG component or a system undergoing some form of interaction, potentially common envelope evolution~\citep{2017A&A...606L...1W}, the latter being a viable solution given the observations.

\noindent\textbf{32~Cyg}\\
32~Cyg has long been a member of the class of VV~Cep type variables and as such undergoes eclipses.
As with all these well-studied targets, the literature is extensive, but some uncertainty exists around the mass of the RSG component in this system. 
\citet{1970VA.....12..147W} prefers a mass measurement of 9.7\,M$_\odot$, but \citet{2007A&A...465..593S} prefer a smaller mass of 7.45\,M$_\odot$.
The orbital period is well known from RV measurements of the RSG following multiple orbits~\citep{2008Obs...128..362G}. 
\citet{2007A&A...465..593S} summarised luminosity measurements for 32~Cyg and we adopt their preferred luminosity.

\noindent\textbf{Antares}\\
Antares is a well known visual binary system, which was first detected in the 1815 lunar occultation and debated throughout the 1800s~\citep{1879Obs.....3...84J}.
While no reliable orbital solution exists for Antares, it is included in this section because estimates exist in the literature of various orbital parameters and, based on its luminosity, Antares A is clearly a high-mass RSG.
\citet{2008A&A...491..229R} determined the orbital period based on the assumption that $e=0$, which they note is unlikely given the observations.
\citet{1978A&A....70..227K} determined the masses of both components via a comparison with evolutionary models for Antares~A, and via a detailed atmospheric analysis for Antares~B.
\citet{2012ApJ...746..154P} determined the luminosity of Antares as $\log L/L_\odot = 4.4$.

This system is clearly distinguished from the others marked in this section, given the size of the orbital period.
Despite the large separation, as with VV~Cep, the hot B-type companion illuminates a nebula that surrounds the two stars~\citep{2008A&A...491..229R}, which was exploited to determine the mass-loss rate of the RSG component in the seminal paper of~\citep{1978A&A....70..227K}.
The origin of the illuminated material is likely the RSG wind.

\noindent\textbf{KQ Pup}\\
\citet{1965ApJ...142..299C} determined the orbital period of KQ~Pup to be 9752\,d with 46 years worth of spectroscopic observations and provides a comprehensive overview of the system.
This star and its binary nature are discussed in detail in \citet{1963PASP...75..509J}.
\citet{1992A&A...256..133R} provide an estimate of the mass of the RSG component of KQ~Pup as 13-20\,M$_\odot$ based on the mass of the companion being 17\,M$_\odot$, which in turn is based on a mass calibration to the assumed spectral type.
We use the term assumed spectral type because, as with VV~Cep, the companion is not directly observed. 
\citet{1992A&A...256..133R} further assumed that, because of the similarity of spectral type to $\alpha$~Ori, these two stars have the same luminosity.
\citet{2024MNRAS.529.3630H} determined the luminosity of KQ~Pup to be $\log L/L_\odot = 4.6$, slightly smaller than the assumed value of 4.8 by \citet{1992A&A...256..133R}.

\noindent\textbf{AZ Cas}\\
\citet{Ashbrook56} is reported to have listed the orbital period of AZ~Cas as 3046\,d by~\citet{1969PASP...81..297C}. 
This orbital period, does not appear to have been refined in the literature since this work.
\citet{2018MNRAS.475.2003D} list this star as having a spectral type of K5\,Iab-Ib.
\citet{1969PASP...81..297C} determined the absolute visual magnitude of AZ~Cas to be $M_V=-4.8$. 
\citet{2024MNRAS.529.3630H} determined a luminosity of AZ~Cas to be $\log L/L_\odot = 4.2$.

\noindent\textbf{5 Lac}\\
\citet{2024MNRAS.529.3630H} determined a luminosity for this source as $\log L/L_\odot = 4.4$.
\citet{2018AJ....155...30B} determined the mass of 5~Lac by measuring the angular diameter distances, which seems to suggest a lower than expected mass of 5.11\,$\pm$\,0.18\,M$_\odot$.
Compared to the other systems in Table~1, the late B-type companion of 5~Lac would seem to support a lower mass solution. 

\noindent\textbf{WY Gem}\\
\citet{1969PASP...81..297C} determined an orbital period of greater than 40\,a. 
Going beyond this, \citet{1998A&A...338..139L} compiled almost 100 years of observations to determine a period of 64\,a with a highly eccentric orbit $e=0.61$. These authors assumed a mass of 20\,M$_\odot$ to determine a mass of the hot companion. 
\citet{1969PASP...81..297C} determined the absolute visual magnitude of WY~Gem to be $M_V=-5.6$.
\citet{2024MNRAS.529.3630H} determined a luminosity for this source as $\log L/L_\odot = 4.6$.

\begin{figure}[ht]
\centering
\includegraphics[width=0.9\linewidth]{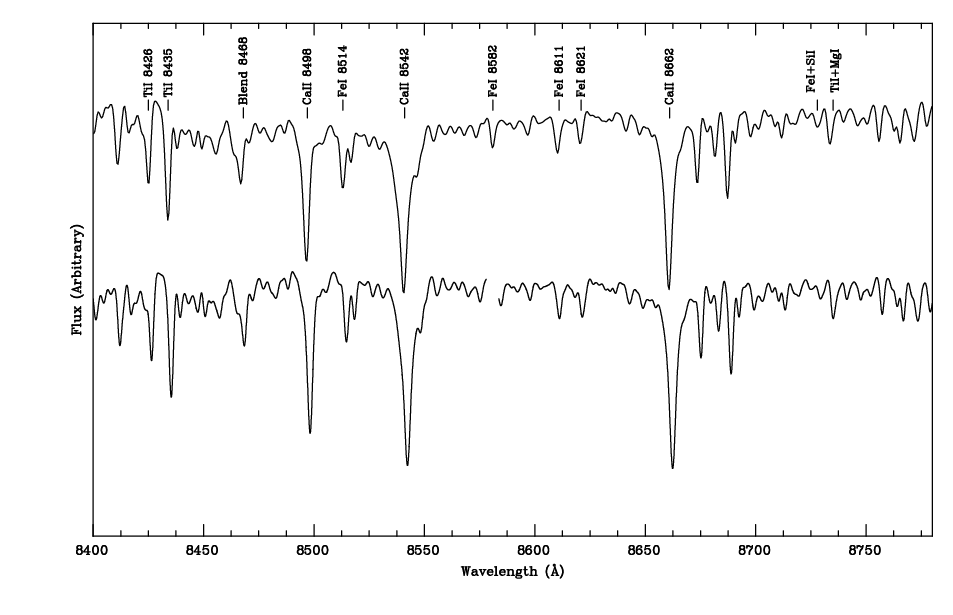}
\bigskip
\begin{minipage}{12cm}
\caption{A portion of the spectrum of TYC~2673-2004-1 (bottom) compared to the spectrum of HD~10\.465, an M2\,Ib supergiant, from the spectral library of \citet{carquillat97}, after convolving to the same resolution. The strength of the Ca\,{\sc ii} triplet and the $\lambda$8468 blend, which are sensitive to luminosity, suggest the same luminosity class. The original spectrum was taken with Mercator/HERMES and has a resolution $R\approx85\,000$ over the whole optical range, except for some small gaps, one of which affects the Fe\,{\sc i}~$\lambda$8582 line.  \label{xsource}}
\end{minipage}
\end{figure}

\noindent\textbf{TYC 2673-2004-1}\\
Confirmed as the optical counterpart of the X-ray source 4U~1954+319 by \citet{masetti06}, TYC~2673-2004-1 was initially classified as an M4\,III giant. The identification was somewhat surprising, as 4U~1954+319, known since the seventies, has always behaved as a typical high-mass X-ray binary. More recently, \citet{hinkle20}, by using the \textit{Gaia} DR2 distance, showed that the star was far too luminous for a red giant. Spectral analysis and existing photometry were employed to determine an effective temperature $T_{\mathrm{eff}}=3450$~K and a luminosity $\log,L/L_{\odot}\approx4.6$. A distance $d\approx3.4$~kpc is confirmed by \textit{Gaia} DR3. A spectrum of TYC~2673-2004-1 in the region of the Ca\,{\sc ii} triplet is shown in Fig.~\ref{xsource}. Based on this, the spectral type would be M2\,Ib. The optical spectrum, as usual for RSGs, gives a slightly later type, close to M3\,Ib. 

The X-ray properties of 4U~1954+319 indicate without doubt that the companion is a neutron star. The X-ray flux shows a strong modulation with a period slightly longer than 5~h. This has been interpreted as the spin period of an extremely slow neutron star \citep{corbet08}. From radial velocity measurements, \citet{hinkle20} constrain the orbit to be rather longer than the $ \approx3$~a necessary to prevent Roche-lobe overflow. The survival of a binary with such a long orbital period after the supernova explosion that created the neutron star puts strong constraints on the presence of a supernova kick, which is also bound to be small because the proper motions of TYC~2673-2004-1 are similar to those of field stars in its surroundings, as was found by \citet{hinkle20} using DR2 and we have confirmed with DR3 data.


Two other systems have been proposed as RSG+NS binaries (excluding Sct~X-1, whose companion seems to be more compatible with a Mira; \citep{de22}), CXO~J174528.79$-$290942.8 \citep{gottlieb20} and  SWIFT~J0850.8$-$4219 \citep{de24,zainab24}. The latter displays emission lines typical of a symbiotic system. The existence of these systems requires supernova explosions with low ejecta masses and very small kick velocities to prevent the disruption of the binary.



\section{Newly discovered populations of RSG binaries in the Milky Clouds}

Multiple, recent, large-scale studies have aimed at identifying and characterising RSG binaries. 
\citet{2024IAUS..361..279P} summarised and compared such results from studies that specifically determined RSG binary fractions in Local Group galaxies. In this article we focus on the observations that characterise the companions of RSGs, which in effect means we focus on the population of UV bright RSG binaries in the SMC~\citep{2022MNRAS.513.5847P} and the HST UV spectroscopic follow-up (presented in Section 4.1). But in this section we first briefly summarise the field more widely.

Using RV measurements of almost 1000 RSGs in the Large and Small Milky Clouds (LMC and SMC, respectively), \citet{2021MNRAS.502.4890D} measured peak-to-peak RVs over a baseline of up to 40\,yrs.
This allowed these authors to identify 45 RSGs (22 in LMC and 23 in SMC) that displayed variability which could only be explained by binary motion. 
Employing the same technique, but covering a baseline of only $\sim$1\,yr,~\citet{patrick2019, 2020A&A...635A..29P} identified binary systems RSGs in clusters in the LMC and SMC, respectively.
The largest peak-to-peak variability found by these authors is consistent with what one would expect from the Figure~1, however, in the MCs, the maximum radius a RSG obtains is smaller because of the reduced metal content in these galaxies, which is expected theoretically~\cite{2011A&A...530A.115B, 2019A&A...625A.132S} and confirmed observationally~\citep{2012AJ....144....2L, 2018MNRAS.476.3106T}.
Studying the orbital motion is important to determine the configurations and allows the identification of the shortest orbital period systems for further follow, but on its own this technique does not reveal the nature of the companions.

\citet{2018AJ....156..225N} studied optical colour-excesses of RSGs and identified binary companions.
\citet{2020ApJ...900..118N} developed this further and used a technique that trained a neural network to identify RSG binary systems based on an observed optical colour excess. 
These authors combined optical and UV photometry from GALEX in the LMC with spectra to train the neural network classifier to identify RSG binary systems in an UBV colour-colour diagram.
\citet{2024arXiv240617177O} applied a similar method of \citet{2018AJ....156..225N} to yellow supergiants in the SMC.
As the photometry and spectra compiled by Neugent et al. are generally in the optical range, contamination from the RSG precludes a determination of the nature of the companion above that the stars have a blue excess.
The spectroscopic confirmations of~\citet{2020ApJ...900..118N} show B-type companions in agreement with observations in the Galaxy, i.e.\ apparently main-sequence early B-type stars.

Using far ultra-violet photometry from the UVIT/Astrosat mission, \citet{2022MNRAS.513.5847P} identified 88 RSGs with a significant UV excess in the SMC.
In the UV survey, the vast majority of the RSGs were not detected.
This, coupled with the accuracy of the UVIT photometry, allowed these authors to conclude that the detections amounted to genuine RSG binary systems with hot companions.
To attempt understanding the underlying distribution of orbital parameters that produced this population of RSG binary systems, \citet{2022MNRAS.513.5847P} used some simplifying assumptions to determine mass-ratios for all their detected systems. 
This led \citet{2022MNRAS.513.5847P} to conclude that RSGs with hot UV companions in the SMC are best described with a flat-$q$ distribution, which was further shown by~\citep{2024IAUS..361..279P}.

This result is perhaps surprising given that these systems have potentially orbital periods up to 10$^8$\,d, which is at odds with other systems with similar orbital periods~\citep{MdS17,Offner_2023}.
However, studies of massive stars in the LMC \cite{2022A&A...665A.148S} and SMC also found a flat $q$ distribution for OB binary systems (see recent results from the Binary at Low Metallicity (BLOeM) campaign~\citep{BLOeM-I} from Sana et al. (\textit{submitted}), Villase{\~n}or et al. (\textit{in prep.})).

Despite their single-epoch photometry, \citet{2022MNRAS.513.5847P}, also attempted to constrain the orbital period distribution of RSG binary systems via a comparison with B-type stars from the LMC~\citep{2015A&A...580A..93D} and concluded that the orbital period distribution drops off sharply after $\log P[{\rm days}] \sim 3.5$.
An examination of Table~1 appears in tension with such a hypothesis, given that only 2/10 of the well characterised RSG binary systems have orbital periods less than $\log P[{\rm days}] = 3.5$.


\subsection{Characterising RSG binaries with HST UV spectra}

Recently, a HST Snapshot survey (PI L. Patrick, PID: 16776) of RSG binary systems was conducted. 
This survey aimed to confirm the nature of highly-likely RSG binary systems in the SMC that displayed a significant UV excess as identified by~\citet{2022MNRAS.513.5847P}. 
This survey obtained HST~STIS UV spectra to characterise the UV spectral appearance and determine more precisely the stellar parameters for the hot companion. 
From the 88 sources identified by~\citet{2022MNRAS.513.5847P}, 28 targets were selected as high-probability RSG binaries, which also included stars which displayed significant RV variability~\citep{2021MNRAS.502.4890D,2020A&A...635A..29P}.
From a total of 28 targets, 17 were observed.
Patrick et al. (\textit{in prep.}) presented these observations and confirmed the presence of the hot star in all cases. In these spectra the contribution from the bright cool supergiant is negligible. 
\begin{figure}
\centering
\includegraphics[width=0.9\linewidth]{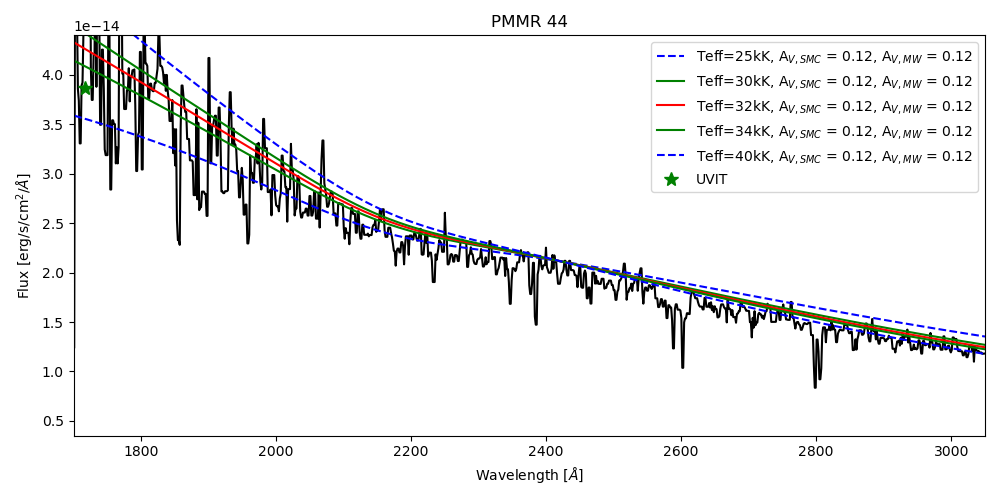}
\bigskip

\begin{minipage}{12cm}
\caption{The full flux-calibrated HST STIS spectrum of PMMR~44 from Patrick et al. (\textit{in prep.}). Coloured lines are effective temperature fits for a fixed A$_V$. The features visible are dominated by interstellar contamination and are not appropriate for stellar parameter determination. The green solid star at 1750\,\AA\ shows the UVIT FUV photometry. The red solid line shows the best fit effective temperature to the continuum for an assumed extinction value, which is significantly larger than the effective temperature determined by~\citep{2022MNRAS.513.5847P}.}
\end{minipage}
\end{figure}

Because of the limitations of their observational data, the stellar parameters determined by \citet{2022MNRAS.513.5847P} relied on three key simplifying assumptions i. the hot companion is a normal, main-sequence, B-type star, ii. the age of the companion is the same as that of the RSG and iii. the extinction parameters are the same for both components. 
Based on Table~1, main-sequence B-type stars appear to be common in such systems, which suggests these assumptions hold for the SMC population, but a more independent determination of stellar parameters is required to assess this.
The HST UV spectroscopic study is able to assess these assumptions by determining stellar parameters largely independently of the cool companion. 
Patrick et al. (\textit{in prep.}) determined stellar parameters for all targets by assuming extinction properties based on the RSG component and compared those to the stellar parameters estimated by~\citet{2022MNRAS.513.5847P}.
In addition, Patrick et al. (\textit{in prep.}) combined \textit{Swift} and UVIT photometry to further characterise the wider population of RSG binaries in the SMC. 
Figure~2 shows an example of continuum fits for a range of effective temperatures for a fixed extinction value to the HST UV spectrum for PMMR~44, a RSG binary system discovered by~\citet{2022MNRAS.513.5847P}.
These authors list the stellar parameters for PMMR~44 as $\log L/L_\odot = 4.43$, $T_{\rm eff}$~=~20\,000\,K.
Based on these data, the effective temperature is significantly underestimated by~\citet{2022MNRAS.513.5847P}, which would result in a more massive companion than expected based on the UV photometry alone.
This is a particularly interesting example as revising up the mass estimate for the companion of PMMR~44, would result in a $q>1.0$ system, assuming the mass of the RSG is accurately determined from the near-IR luminosity.

\section{Summary and future prospects}

In this article, we studied populations of RSG binary systems and concentrate on how the recently discovered systems in the Local Group of galaxies can be understood from a theoretical and observational perspective by considering the best studied RSG binary systems. 
The SMC population of 88 RSG binary systems~\citep{2022MNRAS.513.5847P} is the only extragalactic population where the hot companions can be directly studied.
These systems appear to show a flat $q$ distribution with no overabundance of $q=1$ systems, in slight tension with the population in the Galaxy (although we caution that there are only 5 stars in the latter sample).
Follow-up HST UV spectra will reveal in more detail the nature of the hot companions and is already providing hints that the simplifying assumptions used in the photometric analysis is indeed too simplistic.
Theoretical expectations are discussed in terms of the orbital configurations that are viable for RSGs and their expected companions.
The population of RSG binary systems with well defined orbital solutions summarised and individual systems are discussed.
We focus on systems where orbital periods have been constrained in the literature and, where possible, we report on mass estimates for both components and find a remarkably small number of systems with well defined orbital elements.
In almost all cases of RSGs appearing in binary systems, the companion is a main-sequence B-type star, with the exception of V766~Cen and the small number of systems with a compact companion.

Further searches of Galactic RSG binaries are required to improve statistics and continued observation of the well-studied systems is vital to refine the orbital solution and stellar parameters. 
As many of these systems have accepted orbital parameters based on only one or two orbital cycles and -- in the case for VV~Cep and other enshrouded companions -- the method of tracking the companion is known to be an extremely complicated indicator of orbital motion, it is likely refinements in orbital parameters are required.
For V766~Cen in particular, further observation is of high importance as common envelope evolution is expected to be a relatively short-lived phenomenon and changes in spectral type or nature of the components could occur on short timescales.

For the RSG populations in the MCs~\cite{2020ApJ...900..118N, 2022MNRAS.513.5847P}, multi-epoch spectroscopic information will become available with Gaia~DR4 for a selection of the systems. 
This, combined with longer baseline measurements~\citep[compiled in ][]{2021MNRAS.502.4890D}, will likely allow the characterisation of a significant population of RSG binary systems in the MCs. 
In this context, UV photometry from e.g. Swift and UVIT/Astrosat will be important to constrain orbital solutions and characterise hot companions and identify UV-dark companions.

\begin{acknowledgments}
The authors acknowledge the contribution of D. Lennon, D. Thilker, L. Bianchi, N. Langer, R. Dorda to this project.
L.R.P. acknowledges support by grants
PID2019-105552RB-C41 and PID2022-137779OB-C41 funded by
MCIN/AEI/10.13039/501100011033 by "ERDF A way of making
Europe". I.N. is partially supported by the Spanish Government Ministerio de Ciencia, Innovaci\'on y Universidades and Agencia Estatal de Investigaci\'on (MCIU/AEI/10.130 39/501 100 011 033/FEDER, UE) under grant PID2021-122397NB-C22, and by MCIU with funding from the European Union NextGenerationEU and Generalitat Valenciana in the call Programa de Planes Complementarios de I+D+i (PRTR 2022), project HIAMAS, reference ASFAE/2022/017, and NextGeneration EU/PRTR. \end{acknowledgments}

\begin{furtherinformation}

\begin{orcids}
\orcid{0000-0002-9015-0269}{Lee R.}{Patrick}
\orcid{0000-0003-1952-3680}{Ignacio}{Negueruela}
\end{orcids}

\begin{authorcontributions}
L.R.P. led the conceptualization, data curation, methodology and writing of the text. 
I.N. led data curation for TYC 2673-2004-1 and supported conceptualization, data curation, methodology and writing of text.
\end{authorcontributions}

\begin{conflictsofinterest}
The authors declare no conflict of interest.
\end{conflictsofinterest}

\end{furtherinformation}

\bibliographystyle{bullsrsl-en}
\bibliography{L_PATRICK}

\end{document}